\documentclass[graybox]{svmult}

\usepackage{type1cm}        % activate if the above 3 fonts are
\usepackage{makeidx}         % allows index generation
\usepackage{graphicx}        % standard LaTeX graphics tool
\usepackage{sidecap}          % for side captions with SCfigure command           
\usepackage{multicol}        % used for the two-column index
\usepackage[bottom]{footmisc}% places footnotes at page bottom
\usepackage{url}
\usepackage{newtxtext}       % 
\usepackage{newtxmath}       % selects Times Roman as basic font
\usepackage{tikz}
\usepackage{mathtools}
\usepackage[CJKbookmarks, pdftex, bookmarksnumbered, bookmarksopen, colorlinks, citecolor=blue, linkcolor=blue]{hyperref}
\makeindex             % used for the subject index
\newcommand{\be}{\nopagebreak[3]\begin{equation}}
\newcommand{\ee}{\end{equation}}
\newcommand{\ba}{\nopagebreak[3]\begin{eqnarray}}
\newcommand{\ea}{\end{eqnarray}}

\newcommand{\bc}{\begin{comment}}
\newcommand{\ec}{\end{comment}}

\begin{document}

\title*{Philosophical Foundations of \\ Loop Quantum Gravity}
  \titlerunning{Philosophical Foundations of Loop Quantum Gravity} 
\author{Carlo Rovelli and Francesca Vidotto}
%\authorrunning{} 
\institute{Carlo Rovelli \at Aix-Marseille University, Universit\'e de Toulon, {\em CPT-CNRS}; The Perimeter Institute; and \\ The University of Western Ontario, \emph{Department\,of\,Philosophy} and \emph{Rotman\,Institute\,of\,Philosophy}\\
 \email{crovelli@uwo.ca}\\[1em]
Francesca Vidotto \at The University of Western Ontario\\ \emph{Department\,of\,Physics\,and\,Astronomy, Department\,of\,Philosophy, Rotman\,Institute\,of\,Philosophy}\\
 \email{fvidotto@uwo.ca}\\[1em]
{\it  The University of Western Ontario is located in the traditional lands of Anishinaabek, Haudenosaunee, L\=unaap\'eewak and Attawandaron peoples.}\\[1em]
{\it This is a preprint of the chapter: ``Philosophical Foundations of Loop Quantum Gravity'', to appear in the ``Handbook of Quantum Gravity'', edited by Cosimo Bambi, Leonardo Modesto and Ilya Shapiro, 2023, Springer, reproduced with permission of Springer. }
%The final authenticated version is available online at: http://dx.doi.org/[insert DOI]?.}
}
\maketitle

\abstract{Understanding the quantum aspects of gravity is not only a matter of equations and experiments.  Gravity is intimately connected with the structure of space and time, and understanding quantum gravity requires us to find a conceptual structure appropriate to make sense of the quantum aspects of space and time. In the course of the last decades, an extensive discussion on this problem has led to a clear conceptual picture, that  provides a coherent conceptual foundation of today's Loop Quantum Gravity. We review this foundation, addressing issues such as the sense in which space and time are \emph{emergent}, the notion of \emph{locality}, the role of truncation that enables physical  computations on \emph{finite graphs}, the \emph{problem of time}, and the characterization of the \emph{observable quantities in quantum gravity}.}

%\pagebreak

\section{Introduction}

What is a quantum spacetime?  Are space and time {\em emergent} notions? If so, what do they emerge from, and in which sense?  Is there a fundamental ontology, from which conventional space and time emerge?  Does a quantum theory of gravity require a specific time variable, as  the Schr\"odinger equation does? If not, what is the connection between the common notion of time and quantum gravity? How is evolution described, in the absence of a canonical time or a fixed background spacetime structure? Which empirically observable quantities are well defined in a quantum spacetime? How do we compute their behavior? Are they local is some sense? 

These questions have been discussed widely and at length in the quantum gravity literature, and routinely confuse those entering the field anew.  Here we address them, showing how a coherent {\em conceptual} framework for a quantum theory of gravity can be cleanly defined. 

We give a basic discussion of the notions of `space' and `time'.  This is essential because substantial confusion derives from mixing up different ways in which these notions are used.  We discuss observability in general relativistic physics and in quantum mechanics. Observability in quantum gravity is  subtle precisely because it combines the conceptual subtleties of general relativity with those in quantum mechanics.  We discuss the notion of `emergence' and the so called `problem of time', the precise role of the \emph{finite} combinatorial structures (graphs and two-complexes) that enter concrete calculations, their relation to locality and the reason they are  physically relevant. Our focus is on the conceptual structure of Quantum Gravity in the Loop formulation (LQG) \cite{Rovelli:2004fk,Rovelli2015,Thiemann:2001yy,GambiniPullinBook}. 

\section{Two distinct notions of space}

In its simplest usage, \emph{space} is the structure determined by a relation of contiguity between physical entities.  We use this notion when we say ``I am in London", or `the electron has reached the detector".  In these cases we \emph{spatially locate} entities (our self, London, an electron, a detector) with respect to one another.  This notion of space does not involve metric quantities (distances, areas, volumes...) and refers to a \emph{relation} between entities.  This is a \emph{relational} notion of space. 

In Newtonian physics, a different  notion of space is employed.  ``Space", in this more specific usage, is a \emph{container} in which things, or observations,  are located, it is an entity assumed to exists by itself, independently from the objects or the dynamical degrees of freedom. In this sense, ``space" can also be empty.  Dynamical objects, or observations, are located ``in space", namely they are located with respect to it.  Space, in this sense, can have a geometrical structure, which in Newtonian physics is described by three-dimensional euclidean geometry.
This notion of {\em container} space, formalized by Newton (anticipated by ancient atomism), has played a fundamental role in the development of physics.  We employ it for instance when we use the Newton equations to describe the dynamical of particles as (the evolution in time of) their location in $\mathbb R^3$.\footnote{In the philosophical literature there are  other, distinct, discussions regarding relational aspects of space. One is about the relational aspects of geometry. Another discussion is the confusion between the velocity being relative versus acceleration being relative. A third one regards the possibility of a  relational reading of Newtonian mechanics (in terms of reference systems). Here we are not referring to these issues. The relation we refer to is simple contiguity, which does not require any metric connotation. From the perspective of quantum gravity, Newtonian space is better understood as special configuration of the gravitational field: an entity.  
All these discussions can be traced to the famous distinction at the beginning of the \emph{Principia} \cite{Newton1934}, and to Leibniz's relationalism: ``As for my own opinion, I have said more than once, that I hold space to be something merely relative, as time is, that I hold it to be an order of coexistences, as time is an order of successions". Third Paper, paragraph 4; G VII.363 pg 25–26 \cite{Alexander}). Mixing up these various discussions is an endless source of confusion.}

The distinction between these two notions of space (relational and Newtonian) is essential to get clarity in quantum gravity. In brief: the first extends to the full quantum gravity regime, the second does not \cite{Rovelli2020b}. The reason the relational notion of space survives in quantum gravity is that contiguity --and therefore localization-- can be defined with respect to Newtonian or classical general relativistic spacetime, but can also be defined with respect to other entities, even in the absence of such Newtonian or classical general relativistic spacetime. On the other hand, the reason the Newtonian ``container" notion of space needs to be abandoned, is that Newtonian space is recognised on physical grounds to be an approximate description of a particular configuration of a quantum field. \\   

In its Newtonian version, the second of these notions of space  \emph{emerges}, from the fundamental theory within a series of approximations:  
\begin{enumerate}
\item Newtonian space emerges from Minkowski space in the low relative-velocity approximation (formally, this is a $c\!\to\!\infty$ limit, where $c$ is the speed of light). 
\item Minkowski space emerges from a general relativistic (pseudo-)Riemannian geometry at scales small with respect to the curvature radius (formally, it can be identified with the tangent space at a point). 
\item A (pseudo-)Riemannian geometry emerges from the quantum geometry defined by the LQG states and their dynamics, in a suitable classical limit (formally, this is a $\hbar\!\to\!0$ limit).   
 \end{enumerate}

The relevant notion of ``emergence" here is a weak one, common in physics: {\em in some contexts}, but not always, the physical system admits a convenient and effective {\em approximate} description in terms of an ``emergent" theory. The emergent theory (here: General Relativity) is self-standing and autonomous, and utilizes its own proper notions  (here: relativistic spacetime). These can be related to notions of the underpinning theory: they approximate certain (not all) particular configurations of those.  

Notice that the Newtonian intuition of the existence of space as an entity is not contradicted neither by general relativity nor by LQG. Newtonian space is simply better understood in these theories as the approximate description of a ---classical, or, respectively, quantum--- dynamical field.  With respect to this field, location can be defined relationally, as it can be defined relationally with respect to anything else. 
 
In LQG, the spinnetwork states form a basis of states for the quantum gravitational field. (Later on we shall be more precise about the meaning of  ``state" in this context.) A spinnetwork state $|\Gamma,j_\ell,v_n\rangle$ is determined by a labelled (abstract, not-embedded) graph $\Gamma$ with links $\ell$ labelled by spins $j_\ell$ and nodes $n$ labelled by $SU(2)$ intertwiners $v_n$. These  are a basis of $SU(2)$ invariant tensors in $H_n =\otimes_{\ell\in n}V_{j_\ell}$, where the tensor product is over the links $\ell$ adjacent to the node $n$ and $V_{j}$ is the Hilbert space of the spin-$j$ representation of $SU(2)$.  The nodes of the graph describe elementary quantum excitations or ``elementary quanta" of the gravitational field.  (The interpretation and the physical reasons for these discrete structures is discussed later on, in Section \ref{discreteness}.)

The links of the graph define a notion of contiguity between two nodes linked by the link.  The notion of contiguity between the elementary quanta determine a spatial structure, in the sense of relational space.  Therefore space, in the relational sense, is  present at the foundations of the theory.  In other words, the quanta represented by the nodes of the graph are spatially located with respect to one another. Notice that they are  \emph{not} located --in any sense-- into an external container space.  In a slogan, they are not \emph{in} space; rather, they themselves, with their contiguity relations, \emph{make up} a relational space. 

When matter is present  \cite{Rovelli:2004fk,Rovelli2015,Thiemann:2001yy}, its degrees of freedom are defined as labels on the same graphs $\Gamma$ as the gravitational field states, therefore the notion of contiguity is equally defined with respect to other quanta as well. That is, quanta of space and other elementary quantum field excitations are all located with respect to one another, defining a relational spatial structure. 

Confusion between these two different meanings of ``space" is at the root of a common  misunderstanding of non-perturbative quantum gravity.  Some authors, for instance, suggest  that physics is inconsistent unless it is formulated in terms of observables ``located in space" \cite{Maudlin2007} taking for granted that  a notion of space as a container is necessary for understanding science. This assumption has no base.  

We locate our observations with respect to one another (both spatially and temporally), but the idea that there should be a container space within which observations are located is only a useful theoretical construct well utilized by Newton, not a requirement for intelligibility. For millennia, before Newton, humankind found the world perfectly conceivable in terms of relative localization and not in terms of localization ``in space". In fact, even ``location" in everyday usage usually refers to adjacency to some material object, for instance a location on Earth, and not a location with respect to an abstract unobservable entity such as Newtonian space. 
To claim that science is unintelligible without a container space is to fail to understand the possibility of a  conceptualization of the world different from the Newtonian one. (See also the discussion in \cite{Huggett2013a}),

\section{The emergence of the continuous metric space}

Let us consider the third of the above approximations in some detail. The way a continuous metric geometry emerges from the spinnetwork states is similar to the way a continuous electromagnetic field emerges from the photon states of quantum electrodynamics. A continuous (intrinsic) 3d Riemannian geometry $g$ can be approximated arbitrarily well by a 3d Regge triangulation formed by flat tetrahedra $\tau_n$ connected by triangles $t_\ell$. The (two skeleton of the) dual of this triangulation defines a graph $\Gamma$. Its geometry can be captured by the variables 
\be
G^{\ell\ell'}_{n}=A_\ell A_{\ell'}\ \vec n_\ell\cdot \vec n_{\ell'}
\label{G}
\ee 
where $A_\ell$ are the areas and $\vec n_{\ell}$ the unit normals to the face of the tetrahedron $\tau_n$.  The operators
\be
\hat G^{\ell\ell'}_{n}|\Gamma,j_\ell,v_n\rangle=(\hbar G)^2|\Gamma,j_\ell,\vec E^\ell_{n}\cdot \vec E^{\ell'}_{n}v_n\rangle
\ee 
where the vector operators $\vec E^\ell_{n}$ are $SU(2)$ generators on the $V_{j_\ell}$ tensor component of $H_n$, are defined on the Hilbert space ${\cal H}_\Gamma$ spanned by the spinnetwork states on $\Gamma$.  It can be shown that there are `intrinsic' coherent states $|\psi_g\rangle\in{{{\cal H}_\Gamma}}$ such that 
\be
\langle\psi_g|\hat G^{\ell\ell'}_{n}|\psi_g\rangle=G^{\ell\ell'}_{n}+O(\hbar)
\ee
and the variance of these operators goes to zero with $\hbar$ \cite{Thiemann:2002vj,Livine:2007mr,Bianchi2011}. 

Similarly, the \emph{extrinsic} geometry of a Riemannian 3d space embedded in a 4d spacetime can be approximated by the an extrinsic geometry $k$ of a triangulated 3d space. This is captured by the 4d dihedral angles $\theta_\ell$ between normals to the tetrahedra; ${{\cal H}_\Gamma}$ carries a corresponding operator $\hat \theta_\ell$  \cite{Bianchi:2009ky,Freidel:2010aq} 
and it can be shown that there are (minimally spread) `extrinsic' coherent states $|\psi_{g,k}\rangle\in{{{\cal H}_\Gamma}}$ that satisfy the last equation as well  as 
\be
\langle\psi_{g,k}|\hat\theta_\ell |\psi_{g,k}\rangle=\theta_\ell+\mathcal{O}(\hbar).
\ee
In simpler words: there are quantum states of LQG that approximate Riemannian intrinsic and extrinsic geometries in the classical limit, in the usual sense in which states of any quantum theory approximate configurations of its classical limit.    

Furthermore, there is some evidence (see below) that the LQG transition amplitudes  define a dynamics that approximates pseudo-Riemannian geometries that solve the Einstein equation, arbitrary well.   
In this sense, the pseudo-Riemannian spacetime of classical general relativity can `emerge' from quantum gravity: it is the standard sense in which classical trajectories `emerge' from a quantum theory.  

Various aspects of the Newtonian notion of space are lost when moving to more general frameworks: the special relativistic framework loses the notion of a preferred space foliation which is present in Newtonian spacetime; the general relativistic framework loses the notion of a metric structure independent from the dynamical degrees of freedom; the full quantum gravitational framework lose the notion of continuous physical space.  These notions are useful within approximations, but are not appropriate to describe nature in a full quantum gravitational regime. This is not a problem because they are not needed to have a coherent and intelligible conceptual picture of reality. 

On the other hand, the relational notion of space defined by contiguity of dynamical entities, which is the familiar one we use when talking about space in our everyday life, remains well defined in quantum gravity. 
Hence, certain aspect of the intuitive notion of space (continuity, space as a container, Riemannian geometry...) are \emph{emergent}, others (relational space) are not.  Those that emerge, emerge from the dynamics of the spinnetworks (states of the \emph{quantum}  gravitational field), in a way which is to a large extent similar to the way a continuous electromagnetic field emerge from discrete photons (states of the \emph{quantum} electromagnetic field).

In a full gravitational regime, the metric structure of spacetime displays the typical quantum features.  The most prominent of these are:
\begin{enumerate} 
\item[(i)] The granularity implied by the discrete spectrum of the $\hat G^{\ell\ell'}_n$ operators\cite{Rovelli:1994ge}. This is the most distinct feature and the key result of LQG \cite{Vidotto:2013jia}. (On a possible discreteness of time see \cite{Rovelli2015c} and on a possibility of actually measuring it see \cite{Christodoulou2018b}.)
\item[(ii)] The fact that the geometry can be in quantum superposition of states with definite geometrical properties, with the usual characteristic phenomena such as interference and entanglement. On the possibility of testing this phenomenon (already predicted by non-relativistic quantum gravity), see \cite{Christodoulou2018c}.  
\item[(iii)] The short scale fuzziness due to the fact that the various operators defining geometry do not commute.  Even the $\hat G^{\ell\ell'}_{n}$ operators do not all commute with one another \cite{Ashtekar1998}, hence cannot be diagonalized together: this fact determines the quantum fuzziness of the (intrinsic) 3d geometry at short scale. 
\end{enumerate} 

We close this section addressing two issues raised in the philosophy literature \cite{Huggett2013a}. The first regards the interpretation of the states that are superpositions of spinnetworks with different graphs. How is contiguity well defined if there is more than one graph, and two graphs define a different notion of contiguity? The answer to this question is in the overall structure of the LQG Hilbert space. A Hilbert space ${\cal H}_{\Gamma'}$ spanned by the spinnetworks with given abstract graph $\Gamma'$ is a (proper) subspace of any  Hilbert space ${\cal H}_{\Gamma}$ where $\Gamma'$ is a sub-graph of $\Gamma$. Any specification of a superposition of states with different graphs $\Gamma'$ and $\Gamma''$  must be written as a state in a state space ${\cal H}_{\Gamma}$ where both $\Gamma'$ and $\Gamma''$ are subgraphs of $\Gamma$. (See Section \ref{truncations} below.)  Unless otherwise specified, the quantum superposition between two states with different graphs must be interpret as a state in the Hilbert space ${\cal H}_{\Gamma}$ where $\Gamma$ is the formed by the two \emph{disconnected} components $\Gamma'$ and $\Gamma''$. This resolves any ambiguity.     

The second issue raised in \cite{Huggett2013a} is the observation that the notion of adjacency defined by the graph may not match the one implicitly defined by an averaged large geometry. Hence there may be two distinct notions of contiguity in the theory: the one defined by the graph (that underpins the dynamics of the spin networks) and the one defined by the emergent smooth geometry.  This is correct. The same happens in classical general relativity. A wormhole smaller than the scale of observation can connect macroscopically distant regions of spacetime.  A microscopic notion of contiguity does not need to match the macroscopic one.  There can be ``wild" geometries in classical general relativity, where a similar mismatch happens, and this does not jeopardize intelligibility.\footnote{In addition, regular geometries can be discretized in terms of ``wild" triangulations. See later.} Similarly, there are ``wild" states in LQG. We do not know the physical relevance (if any) of either. 

To clarify why two notions of adjacency do not represent a problem, it is useful to ask what is the physical meaning of adjacency.  The answer should not be searched in a Kantian {\em a priori} condition for intelligibility, but in what we have learned from experience about the world around us. The adjacency  relation that we experience is rooted in the fact that physical interactions are local. Because of this, we only directly affect --and we only directly receive information from-- adjacent entities.  In other words, the basic spatial structure of the world is determined by what directly affects what.  
The dynamics of loop quantum gravity is local on the graph (both in the Hamiltonian and in the covariant formulations of LQG). This is why the locality relation that we experience derives from the locality defined by the graph structure of the states.

\section{Observability in gravitational physics}
\label{grob}

The conceptual structure of general relativity (GR) is subtle and has confused all relativists (including Einstein) for a long time.  For this reason, decades have been necessary before getting clarity on question like the nature of the Schwarzschild singularity, or whether gravitational waves are physical or gauge artifacts.  

Much of the confusion stems from the fact that the theory is written in terms of spacetime coordinates $x$ and $t$, but the  physical meaning of these is totally different from the physical meaning of the spacetime coordinates with the same name used in special relativity and in non-relativistic physics.  The spacetime coordinates $X$ and $T$ in non relativistic and special relativistic physics have \emph{metric} meaning: the spacetime coordinates $x$ and $t$ in general relativistic physics do not have metric meaning. 
That is, in special relativity for a particle to have position $X$ means  to be at a physical distance $X$ from the axis of some established physical reference frame. This distance can be measured with a rod, a laser, or anything else. For an event to have coordinate $T$ means to happen when a clock has measured a time lapse $T$.  Not so in GR: in GR, if a particle has position $x$ this does not mean that the particle is at a physical distance $x$ from something. For an event to have coordinate $t$ does not mean that a clock has measured a time lapse $t$ from some initial time. Distance (measured in any of the above manners) and clock readings are rather given by integrals involving the gravitational field, such as 
\be
T=\int_\gamma \sqrt{g_{ab}dx^adx^b}, 
\ee
where $g_{ab}(x)$ is the gravitational field. The value of these integrals does not change if the coordinates are changed to new coordinates.   The fact that the coordinates have such a dramatically different meaning in the two contexts raises continuous confusion. 

Related to this is a persistent confusion about the connection between the theory and the physical measurable quantities, usually called ``observables". The reason of the confusion is the following. The Einstein equations are invariant under arbitrary changes of the  coordinates $x$ and $t$. It follows that in general a quantity that depends on the coordinates $x$ and $t$ cannot be predicted by the theory.  

There are three alternative ways of interpreting this fact and using the theory, all three equally valid (see \cite{Rovelli:1990ph}). 
\begin{enumerate}
\item The first is to only consider observables that are invariant under coordinate transformations. These are predicted by the Einstein equations, once a solution is specified. For instance, the minimal distance between the Earth and the Moon during the current month, as measured by the return (proper-)time on Earth of a laser signal bounced off the moon, is a quantity that does not depend on the coordinates chosen. All quantities measured in relativistic observational gravity can be interpreted in this manner. 

\item The second option is to interpret the coordinates as labels of concrete reference objects whose dynamics is determined by the theory. This is a gauge fixing of the coordinate choice, and as such it promotes the coordinates to quantities that can be actually determined physically.   This procedure is commonly followed for instance in cosmology (in the homogeneous approximation), where coordinates are attached to galaxies and the proper time on these.  In this language, the Einstein equations are gauge fixed on particular coordinate choices. 

\item The third option is again to interpret the coordinates as labels of concrete reference objects but to disregard the dynamical laws governing these reference objects. The under-determination in the evolution equations can then be interpreted as the result of disregarding these dynamical laws, namely choosing physical reference systems that move arbitrarily in spacetime. 
\end{enumerate}

The three options are all viable, and ultimately equivalent. They refer to different sets of variables. While the first and the second refer to the physics of the dynamical degrees of freedom included in the theory, with nothing else interacting, the third refers to the the physics of the dynamical degrees of freedom of the theory {\em interacting with other degrees of freedom}.  

In other words, the gauge degrees of freedom of general relativity can alternatively be:
\begin{enumerate}
    \item interpreted as unphysical, namely just as a redundancy of the mathematics; or 
    \item gauge fixed; or
    \item interpreted as relational degrees of freedom describing the coupling with an external (arbitrarily moving) physical system, used as physical reference system \cite{Rovelli2014a}.
\end{enumerate}
In all interpretations, spacetime localization is only relative.  In the first case, objects and events in the theory are localized with respect to one another (the Earth and the laser pulse). In the second, they are localized with respect to the chosen reference system (for instance, the galaxies and their clocks).  In the third, they are localized with respect to the external arbitrarily moving reference system. 
Much of the conceptual confusion about the the observables of general relativity and about the interpretation of "spacetime points" comes from mixing up these three cases.  

A consideration is important for what follows.  Are observables in general relativity  \emph{local} in some sense?  Let us consider two examples taken from actual applications of the theory, where physical quantities are concretely measured by experimentalists and astronomers, and compared to the theory.  As a first example, consider a detection of a gravitational wave pulse by a gravitational interferometer. This can be thought as a curvature measurement in a location defined by components of the detector.  It is  \emph{local} in the sense that it only involves what happens in the location of the detector. A second example is a typical measurements in the analysis of the general relativistic dynamics of the Solar System.  Astronomers measure the physical distance between a given point on Earth and a given point on Venus, at a specific time determined by some event of Earth, and defined as half the forward-backward travel time of a laser pulse that bounces off Venus, where the travel time is in terms of the Earth proper time.  The measured quantity is  \emph{local} in the sense that it only involves what happens in the Solar System.   Now, say we interpret the theory according to the first option above.   Are the actual measured diffeomorphism  invariant  variables   \emph{local}, in the sense that they can be expressed as local functions of the coordinates in the dynamical system formed by all the entities involved?  Of course they are not, because no diffeomorphism invariant quantity is local in this sense.  These examples should lead us to caution: one often reads that the absence of ``local'' observables represents a major obstacle in interpreting general relativity and quantum gravity.  This is certainly not the case in the classical theory, as shown by these examples.  We shall come back on this in the quantum context. 

Notice that if we adopt the third reading of general covariance, the interpretation of what is measured in the two examples above simplifies dramatically.    Take the case of the detection of the gravitational wave pulse.  In the first interpretation, we consider the coupled dynamical system formed by the gravitational field and the interferometer, and the measured quantities is a highly non local function of the basic variables.  In the third interpretation, instead, we can think that the system under consideration is just the gravitational field, and view the laboratory containing the detector as an external (``reference'') system with which the gravitational field is interacting.  The quantity measured by the detector is a \emph{local} function of the metric, in the location determined by the detector.   The full diffeomorphism invariance of the pure gravity dynamics, in other words, is physically broken by the detector itself being located somewhere.   

The two interpretations are equally correct, and both have advantages. As we shall see, the second opens up an interesting window of opportunity in the quantum context, in relation to the necessity of considering a Heisenberg cut in quantum measurements.

Mixing up three interpretations above is also the source of the confusion in the discussion about the meaning of spacetime points in general relativity (and the ``hole argument" \cite{Norton2019}) which has been going on in philosophy of science \cite{Roberts2020}.  The discussion is confused by the fact that in the first interpretation there is no physical definition of points independently from the degrees of freedom \emph{of the theory}, but in the third there is such definition (because the points, individuated by the coordinates, are defined relationally with respect to the external arbitrarily moving reference system.)

The third of the above interpretations is the reason for the strong (irresistible, for some) intuitive appeal of the reality of a manifold independent from the value of the gravitational field defined over it.  The points of the manifolds are possibilities for coupling other degrees of freedom.  Once we include all degrees of freedom, the manifold is dispensable (as many relativists like to repeat \cite{Earman}).  The same is true for the graph of a spinnetwork and the two-complex of a spinfoam, if these are considered independently from their labeling. 

This discussion, more broadly, also sheds light on the general interpretation of gauge invariance: gauge is more than mathematical redundancy, because the gauge degrees of freedom capture ways a physical system can couple with other physical systems. This is because (gauge invariant) couplings can couple to gauge variant variables of a component-system. A discussion on this fact is in \cite{Rovelli2014a}.

\section{General relativistic evolution}
\label{evoluzione}

Physics describes processes, namely how things happen, or how they ``change". To do so, general relativistic physics employs a more subtle notion of evolution than Newtonian physics. 

In Newtonian physics, evolution is described by writing equations that govern how physical variables change \emph{in time}.  In general relativistic physics, dynamical processes are described by writing equations that govern how physical variables (including those characterising clocks) change \emph{with respect to one another} \cite{Rovelli1994b,GambiniPullin,Giesel:2006uj,Domagala:2010bm}.  A characteristic example of utilization of this relative notion of evolution is Loop Quantum Cosmology  \cite{Agullo2013a} where the dynamics of the universe is often coded in the relative evolution between the cosmological scale factor and the value of a homogeneous scalar field. 

More precisely, in Newtonian physics we use dynamical variables  $A, B,...$ plus a `special' (preferred, canonical,...) time variable $T$.  We call `clocks' the measuring devices that best track this variable. The time variable $T$ is used as the independent variable of the evolution, and we write equations of motion for the functions $A(T), B(T),...$.   A motion can equally be represented as a line in the space of the variables $(T, A, B,...)$, defined implicitly (`covariantly') by functions of the form $f(T, A, B,...)=0$.

In a general relativistic physics, evolution is described by writing equations that govern how physical variables (including those measured by clocks) change \emph{with respect to one another}.  This is because there is no single canonical time variable. Different `clocks' determine distinct measurable variables $T_n, n=1,2,...$.   Accordingly, we define evolution by giving relations between all variables including the clocks. A motion is therefore described by a line in the space spanned by all variables $(T_n, A, B,..).$, defined covariantly by functions of the form $f(T_n, A, B,...)=0$.  This line can be parametrized by an arbitrary label $t$: this is the relativistic time coordinate, which should not be confused with the readings $T_n$ of clocks. 

The above is easily generalized to field theory.  In a 4d (non-general-relativistic) field theory, evolution is  described by equations for fields $A(X,T), B(X,T),...$ that depend on spacetime coordinates $(X,T)$. These coordinates represent distances measured by rods and time intervals measured by clocks.  A motion can equally be represented as a 4d surface in the space spanned by the variables $(X, T, A, B, ...)$. In a general relativistic field theory like general relativity, on the other hand, evolution is described by writing equations that govern how physical variables (including distances measured by rods and time intervals measured by clocks) change \emph{with respect to one another}.  Evolution is given by relations between all variables including clock variables $T_n$ and distance variable $X_n$. A motion is a 4d surface in the space of all variables $(X_n, T_n, A, B,...)$.  This surface can be parametrized by arbitrary 4d labels $(x,t)$: these are the relativistic spacetime coordinates. 

The fact that there is no preferred time variable in relativistic gravitational physics can be seen for instance by noticing that in general two clocks measure different (\emph{proper}) times between the same couple of events, depending on their location, speed, etcetera.   Consider the following example:  launch a clock $C_1$ upward at some moment and catch it back when it falls at a second moment. In the meanwhile,  hold a second clock in your hands. The two clocks will measure two different times $T_1$ and $T_2<T_1$ between the launch and the catch. Which one is \emph{the real} time variable?  The answer is that there is no ``real" time : both times are physical times, and can be taken as independent variables. 

The way evolution is treated in general relativistic physics is reflected in the Hamiltonian structure of the dynamical theory.   The general structure of Newtonian physics is given by a Hamiltonian $H$ on a phase space $\Gamma$. The phase space is a symplectic manifold.  In a symplectic manifold, a function $H$ generates a flow: the physical motions are the orbits generated by the flow of $H$  in $\Gamma$.

General relativistic physics requires a generalization of this structure, which we sketch here (for more details, see \cite{Rovelli:2004fk}). The generalization is given by a constraint $C$ on an extended (symplectic) phase space $\Gamma_{ex}$.  The symplectic form on $\Gamma_{ex}$ induces a pre-symplectic form on the constraints surface $C=0$. The motions are the lines (surfaces in field theory) on the constraint surfaces whose tangents are null directions of the pre-symplectic form.  The general (finite dimensional) case reduces to the Newtonian case when $\Gamma_{ex}$ is the Cartesian product of $\Gamma$ and a space with canonically conjugate coordinates $(T,p_T)$, and $C=H+p_T$, as can be easily verified.  

The quantities like $T, A, B,...$, that include dependent as well as independent dynamical variables are called \emph{partial observables} \cite{Rovelli2002b}.  These are quantities that can be measured but cannot be individually predicted even with full knowledge of the motion (as they include the independent variables of the evolution). What the theory predicts is not the value of individual partial observables, but, rather, relations among them. For instance, for a harmonic oscillator with a single degree of freedom, its position $X$ and the time $T$ are both partial observables, and once the motion is known (amplitude and phase are known), the theory predicts the value of $X$ for any given $T$, or the possible values of $T$ for any given $X$. 

The \emph{physical phase space} is the space of the motions. It is again a symplectic space, and in the case of a Newtonian theory is isomorphic to the familiar space of the initial data, but not canonically so, because the isomorphism is determined by a value of $T$.  Each point of the physical phase space determines a relation between partial observables. 

The motions can be parametrized.  That is, the surfaces defined by the functions $f(A_n)$, where $A_n$ coordinatize the space of the partial observables of the theory, can be written as functions $A_n(x)$ of arbitrary coordinates $x$. The spacetime coordinates used in general relativity are these parameters. 

In quantum theory, the partial observables are represented by self adjoint operators on an \emph{extended} Hilbert space ${\cal H}_{ex}$ and the dynamics is given by a constraint operator (or set of operators) $C$ defined on ${\cal H}_{ex}$.  The transition amplitudes that define the quantum dynamics amplitudes are given by 
\be
W(a,b)=\langle a |P | b \rangle
\label{W}
\ee
where $ | a \rangle$ and $ | b \rangle$ are eigenstates of (a complete set of commuting) observables defined on ${\cal H}_{ex}$.  $P$ is the projector of the kernel of $C$ if this is a proper subspace of ${\cal H}_{ex}$.  If zero is in the continuum spectrum, $P$ can equally be defined using distributional techniques, see for instance \cite{Marolf1994,Marolf1995,Rovelli:2004fk}.  

The transition amplitudes \eqref{W} define the quantum dynamics of a general covariant quantum field theory. 

To compute probabilities from the amplitudes we must remember that  $H_{ex}$ describe partial observables, which include the independent variables. We can therefore only assign probabilities to some components (say, $a_1$) of $a=(a_1,a_0)$ at given value of others ($a_0$). That is, probabilities are well defined when
\be
\sum_{a_1} |W(a_1,a_0,b)|^2=1. 
\ee
If the set of variables $a_o$ include a variable $t$ such that the dynamics is symmetric under a translation in $t$, then the Hilbert space carries a unitary representation of the group $\mathbb{R} $. If there is no variable with this property, then there is no unitarity in this sense in the theory, but this does not mean that probabilities are ill defined.  For an enlightening simple example of a well defined quantum system without unitarity in this sense, see \cite{Colosi:2003si}.  For a discussion on the definition of probability in the general case, see \cite{Oeckl:2005bv}.

\section{Observability in quantum physics}

Classical physics assumes that all properties of a system are always sharply defined.  Not so quantum physics. Properties are given by eigenvalues of observables and quantum mechanics only assigns properties to a system in the context of an interaction with another system. The boundary between the two is called the \emph{Heisenberg cut} \cite{VonNeumann1955}.    For instance, in the Copenhagen interpretation properties are actualized in measurements; in the relational interpretation, they are always relative to a second system; in the many-world interpretations, they depend on the branching and are related to Everett's relative states determined by a split due to a Heisenberg cut.\footnote{The terminology ``Heisenberg cut" is characteristic of the Copenhagen interpretation, but we use it more generally to denote the separation between systems which is needed in order to have actual values of variable also in interpretations such as Many Worlds and Relational.}  

For simplicity, we use the language of the Copenhagen interpretation, but what we say can be easily translated in the language of different interpretations.  In the Copenhagen interpretation we mentally distinguish the system from the context, separated by the Heisenberg cut and treat the context classically.  The observables of the system take a value at a measurement, which is an interaction between the system and the context.  The theory predicts the probability of one or the other value to be actualized in this interaction (called \emph{measurement}), given that other values of observables were actualized in a previous interaction (called \emph{preparation}).  The cut can be moved outward without changing the predictions. Denoting $a$ the set of the observables' values actualized in the preparation and $b$ the set of the observables' values actualized in the measurement, the conditional probabilities predicted by the theory are given by
\be
P(b|a)=|W(b,a)|^2
\ee
where the transition amplitude is given by
\be
W(b;a)=\langle b|a \rangle;
\ee
here $ | a \rangle$ and $ | b \rangle$ are the relevant normalized eigenstates of the operators corresponding to the relevant observables. (In the relational interpretation the cut defines relational observables; in the Many World interpretation, the cut separates two subsystems that defines the branching, withing which variables have determined values...)

If a time $(t'-t)$ different from zero lapses between the two measurements and the hamiltonian is $H$, the transition probabilities are
\be
W(b,t';a,t)=\langle b|e^{-\frac{i}{\hbar}(t'-t)H} |a \rangle;
\ee

In the general relativistic case, the time variable is included among the partial observables, and we write the amplitude above as  
\be
W(b,t';a,t)=\langle b,t'|P|a,t \rangle.
\ee
Additionally, we can consider a `boundary' Hilbert space $H_{b}=H_{in}\otimes H_{out}$ namely the tensor product of the in and out state spaces, and express the dynamics as a single (possibly generalised) bra on this boundary state space
\be
W(b,t';a,t)=\langle W | b,t';a,t \rangle.
\ee
For details on the formalism, see \cite{Rovelli:2004fk}.

\section{Observability in quantum gravity}
\label{observability}

To understand observability in quantum gravity we have to combine our understanding of observability in general relativity with our understanding of observability in quantum theory.  From the second, we learn that 
 what the theory can predict is the probability of one or the other property of the gravitational field to be actualized in interactions, across a Heisenberg cut, with a context that can be treated as classical. The natural and simple key for this to work is to identify the Heisenberg cut with the boundary of a four dimensional spacetime region $\cal R$ \cite{Conrady2003b}.  It is not difficult to see that any realistic observation in relativistic gravitation can be expressed in this form. 

It is particularly convenient to take $\cal R$ to be compact, and bounded by a 3d surface $\Sigma$ formed by the union of a past and a future spacelike surfaces $\Sigma_-$ and $\Sigma_+$ joined along a two sphere.  
The theory is naturally expressed in the {\em time gauge} on these surfaces. 
This setting permits us to interpret the LQG transition amplitudes as transition amplitudes from  $\Sigma_-$ to $\Sigma_+$.  Quantum states on  $\Sigma_-$ and $\Sigma_+$ represent quantum geometries on these surfaces.  These are interpreted as interactions between the gravitational field on $\cal R$ and the rest of the universe, across the Heisenberg cut defined by $\Sigma$. 

This setting works very well in the two contexts where we expect quantum gravitational phenomena to be non negligible: early cosmology and around black hole singularities. (See below.)

A preparation and (a complete) measurement at the Heisenberg cut, namely at $\Sigma$, determines the eigenvalues of a complete set of (\emph{partial}) observables, and a corresponding  eigenstate $|\Psi\rangle\in{\cal H}_{b}$. (Semiclassical coherent stares are more convenient in some applications.)  The dynamics is then  given by a single bra $\langle W|$ on ${\cal H}_{b}$. This is the covariant version of the transition amplitude 
\be
\langle W|\psi\rangle = \langle \Psi_+|P|\Psi_-\rangle 
\ee
where $P$ is discussed in Section \ref{evoluzione}, the boundary Hilbert space is discussed in the last Section and $|\Psi_\pm\rangle$ are quantum states of the geometry of, respectively, $\Sigma_\pm$.  

Formally, if the measured quantities correspond to the 3d (intrinsic) geometry $g$ on $\Sigma$ and we call the corresponding eigenstate $\Psi_g$ we may write 
\be
\langle W|\Psi_g\rangle = \int_{\partial g_4 = g} Dg_4\ e^{-\frac{i}{\hbar}\int \sqrt{-g_4}R[g_4]},
\label{fi}
\ee
where the (ill defined) functional integration of the exponent of the Einstein Hilbert action is over the 4d geometries $g_4$ on $\cal R$ bounded by $g$ on $\Sigma$, as originally suggested by John Wheeler and  Charles Misner \cite{Misner:1957fk}.  The spinfoam formalism can be viewed as a way to transform the last formula into something well defined and computable, within arbitrary truncations. 

In a semi-classical regime we expect  the classical dynamics of general relativity to be recovered from the approximation 
\be
\langle W|\Psi_g\rangle \sim \sum_{n} e^{-\frac{i}{\hbar}S_n[g]},
\ee
where the sum is over the different solutions $g_4[g]$ of Einstein equations on $\cal R$ that induce the 3-metric $g$ on $\Sigma$ and 
\be
S_n[g]=\int_{\cal R}\sqrt{-g_4[g]}R[g_4[g]]
\ee
is the corresponding Hamilton function of general relativity \cite{Rovelli:2004fk}, namely the value of the action on a solution with given boundary data.  As well known, full knowledge of the Hamilton function is essentially equivalent to knowledge of the solution of the equations of motion. (To see how this still works in the generally covariant case, see \cite{Rovelli:2004fk}.)

From the conceptual point of view that concerns us here, we notice that the observables of quantum gravity can be chosen to be sitting on the Heisenberg cut $\Sigma$ and be partial observables. Importantly, they do not need to be fully gauge invariant, because they represent interactions between the quantum system studied and (`the measurement apparatus on') the boundary of the classical context.  What the theory provides, then, are transition amplitudes between partial observables, and these give the physical predictions, as illustrated above. 

An alternative way to derive gauge invariant predictions from quantum gravity is to write fully gauge invariant observables on the phase space of the theory. If we had a sufficiently rich family of such gauge invariant observables, these could be used to  compute transition amplitudes directly, because fully gauge invariant observables in general relativity are like operators in the Heisenberg picture. The scalar product between their eigenstates would directly give transition amplitudes, exhausting all relevant dynamical information.  Many authors have explored this path  \cite{Bergmann:1960wb,Bergmann:1961wa, Page:1983uc,Rovelli1991, Perez:2001gj, Rovelli:2001my, Dittrich:2004cb, Dittrich:2005kc, Dittrich:2006ee,Giddings2006,Giesel:2009jp,Kaminski:2009qb, Donnelly2016a,Donnelly2016b, Duch2014,Duch2015,Bodendorfer2015,Duch2016}. This strategy involves an infamously hard task, as discussed in detail by many of the authors cited, from a variety of points of view: Dirac observables are highly non-local and obey a non-local algebra in a  general relativistic setting.   The difficulty has been addressed in various manners, such as expanding around flat space, or coupling suitable matter fields to use as references, or using nonlocal dressings of fields, or using geodesics from infinity to define bulk localization, or considering the particular class of Dirac observables called the `evolving constants of the motion' \cite{RovelliQM3}. Here we shall not review those intriguing possibilities, for which we refer to the authors cited. 

The  boundary strategy for computing predictions in quantum gravity described above, instead, circumvents the infamous difficulty of writing fully gauge invariant observables (or ``Dirac observable'') in general relativity as explicit functions of the theory's phase space. {Explicit knowledge of these functions is not necessary to extract information from a quantum theory of gravity}.  In fact, it is not necessary to extract information from the {\em classical} theory of gravity either. This is evident form the fact that we do not know how to write such quantities on the phase space of general relativity, and yet in general relativistic physics, astrophysics, and cosmology, we can routinely do general relativistic observations, measurements, and predictions, as illustrated in Section \ref{grob}.  Clearly there are ways of extracting physically meaningful information from a general relativistic theory even without being capable of writing its Dirac observables.  

In other words, to make predictions about the behaviour of a covariant system it is not necessary to know explicit its physical Hilbert space and the operators which leave the physical Hilbert space invariant. Knowledge of the extended (or kinematical) Hilbert space ${\cal H}_{ex}$ and the transition amplitudes is sufficient.  For example, consider a particle in two dimensions described in a covariant manner with states $|x,t\rangle\in{\cal H}_{ex}$. The full theory can be expressed in terms of the transition amplitudes
\be
   W(x,t;x',t')\equiv \langle x,t|P|x',t'\rangle, 
   \label{WW}
\ee
where $P$ is an operator on ${\cal H}_{ex}$, without reference to a physical Hilbert space of solutions of the ``Wheeler-deWitt equation'' whose solutions define the physical Hilbert space.  If we can deparametrize the theory, the explicit form of the transition functions can be written in the well known form
\be
   W(x,t;x',t') \equiv \langle x|e^{-\frac{i}{\hbar}H(t-t')}|x'\rangle, 
\ee
but this expression may be ill defined in the general case, while \eqref{WW} remains well defined.   Intuitively, and in the cases where these equations are well defined,
\be
   W(x,t;x',t')\sim \langle x,t|\delta(C)|x',t'\rangle \sim \int_{(x',t')\to(x,t)} DX \ e^{{\frac{i}{\hbar}S}[X]}
   \label{W}
\ee
where $C$ is the Hamiltonian constraint, $S$ the action and the integral is a Feynman integral over paths $X$.  In the covariant formulation of LQG, the (truncated, see below) quantities $W$ are defined by the spinfoam amplitude and are functions of boundary states representing 3-geometries, as in \eqref{fi}. There is no need to "deparametrize" or finding Dirac observables, to study quantum gravitational processes. 
Quantum transition amplitudes, like predictions in classical general relativity, can be formulated and computed using only partial observables, working with gauge dependent quantities and exploiting the third of the three interpretations of general covariance listed in the Section \ref{grob}.

The relational interpretation of quantum theory \cite{Rovelli:1995fv,Laudisa2017} is a natural setting when quantum gravity is formulated in this manner.  The relational structure of space and time merge naturally and beautifully with the relational structure of quantum theory: the Heisenberg cut is identified with spacetime partitions. Notice that in the Copenhagen version, we still need an outside classical spacetime. Not so in the relational interpretation, where what matters in only that there is am (relational) boundary between two systems, without any presupposed geometry on this boundary. 

However, the problem of quantum gravity and the problem of the interpretation of quantum mechanics are distinct and to a large extent independent. 

A concrete example of utilization of LQG transition amplitudes is given by the calculations of what happens at the end of the evaporation of a black hole.  Most of spacetime can be treated classically, because quantum gravitational effects are negligible.  Not so the high curvature region surrounding the classical (unphysical, because of quantum gravity) singularity and the horizon near the end of the evaporation.   This is the compact quantum region $\cal R$. A 3d surface surrounding it can be chosen and the LQG quantum transition amplitudes describing what can happen at the end of the evaporation can be explicitly studied: see \cite{Bianchi2018c,Rovelli2014ps,DAmbrosio2021,Christodoulou2018d,Soltani2021}.     (This is possible because a \emph{classical} exact solution for the exterior exists \cite{Haggard2014}.)

Another concrete example is the use of this covariant formalism to study the big bang \cite{Bianchi:2010zs,Gozzini2019}. In this case the surface $\Sigma$ can be taken to be a single spacelike surface with the topology of a 3-sphere after the big bang, to describe the transition from nothing to a cosmological space (as in the Hartle-Hawking scenario), or, alternatively, as two disconnected spacelike surfaces with the topology of a 3-sphere, to describe a Big Bounce, as in Loop Quantum Cosmology  \cite{Vidotto:2011qa}. 
In this cosmological context, the average value of the spins can be taken as the independent variable (a discretized version of the cosmological scale factor), in terms of which the dynamics of the fluctuations of the rest of the geometry evolves.

\section{Truncation, finite graphs and finite spinfoams}
\label{truncations}

The bra $\langle W|$ that gives the dynamics is defined in covariant LQG  in terms of spinfoam amplitudes,  order by order in a suitable sequence of truncations that represent increasingly fine approximations. See \cite{Rovelli2015} for a detailed technical introduction of these.   Here we only discuss the conceptual structure of the theory. 

A general remark is important here, to dispel a recurring conceptual confusion: the idea that a quantum theory, and in particular a quantum theory of gravity, could only describe \emph{elementary} components of nature.  There is no reason to expect so, and this is not the right way of viewing and using quantum theory.  Quantum theory is not a theory about the elementary components of reality: it is a theory about the quantum behaviour of \emph{any} physical variable, irrespectively on whether this is elementary or composite. The angular momentum of a molecule, for instance, is quantized and can be used in computing transition amplitudes in the dynamics of the molecule, independently from the internal quarks's structure of the molecule. 

In the same manner, there is no reason to see the quantization of some aspects of the metric field as a description of elementary components: any variable, at any scale, behaves quantum mechanically and LQG is the description of quantum properties of gravitational degrees of freedom, {\em at any relevant scale}.  A state does not represent a ``thing'': it represents  the outcome of an interaction in which certain observables take on certain values.  No measurement delivers an infinite amount of information. Any measurement captures a \emph{finite} number of degrees of freedom only.  {\em An effective theoretical description of a given phenomenon needs only to refer to the degrees of freedom that are relevant for that phenomenon}. 

This is also true in quantum field theory.  In standard quantum field theory calculations are always performed on finite lattices and finite graphs.  For instance, in lattice QCD hadron's masses are computed on lattices of \emph{finite} size that are large enough to include the hadron and fine enough to see the quarks' wavelengths, but no more. Analogously, collisions in QED and in the electroweak theory are computed order by order in a perturbation expansion: at each order there is a maximum number of (real and virtual) particles involved, and therefore, again, only a finite number of degrees of freedom involved. For this reason, both the lattices concretely used in QFT calculations and the (Feynman) graphs in QFT perturbation theory are finite (that is, they have a finite number of vertices). 

The same holds in loop quantum gravity: concrete calculations  involve spin networks and spinfoams with finite graphs and finite two-complexes. The calculation is expected to provide results that approximate the physical behavior of phenomena where the corresponding degrees of freedom play a role. 
When describing a phenomenon (like the end of the evaporation of a black hole, or the possible bounce of the primordial universe), we have to single out the degrees of freedom that may play a relevant role in the corresponding dynamics, and describe the process in terms of these, not in terms of everything \cite{Borja:2011di,Vidotto2017}. 

Suggestions that calculations on finite graphs and finite spinfoams  are unreliable are therefore conceptually ill-founded. 

A measurement of the geometry that captures a finite number of degrees of freedom can be modelled as follows.  Given a 3d metric space with geometry $g$, consider a simplicial decomposition of the space and call $A_\ell$ the areas of the 2-simplices $\ell$, and $\vec n_{\ell}\cdot \vec n_{\ell'}$ the angle between two vectors normal to two 2-simplices $\ell$, and $\ell'$ bounding the same 3-simplex $n$, in an arbitrary point of $\ell$ and $\ell'$, parallel transported to an arbitrary internal point of $n$. Then equation \eqref{G} defines a family of quantities that measures the geometry $g$ at some scale.  These quantities do not commute in quantum theory. 
A (smaller) set of commuting quantities is given by the areas $A_\ell$ of the two simplices and the volumes $v_n$ of the three simplices.   A volume operator $V_n$ is defined on ${{\cal H}_\Gamma}$, where $\Gamma$ is the graph dual to the cellular decomposition. The operators $(A_\ell,V_n)$ form a commuting set of operators in ${{\cal H}_\Gamma}$ which is maximal up to some signs that we disregard here for simplicity.  These operators have discrete spectrum  \cite{Rovelli:1994ge}.  Let $|\Gamma, j_\ell, v_n\rangle$ be a basis in ${{\cal H}_\Gamma}$ that diagonalizes this set. The states $|\Gamma, j_\ell, v_n\rangle$ can be interpreted to represent the outcomes 
$(j_\ell, v_n)$ of these measurements, in the standard sense of quantum theory.

This does \emph{not} mean that a state like $|\Gamma, j_\ell, v_n\rangle$ gives a complete description of reality in a certain spacetime region. It \emph{only} refers to a subset of degrees of freedom measured.  The theory that describe \emph{these} degrees of freedom is a good description of reality to the extent the dynamics of these degrees of freedom is not too affected by others. 

In the covariant formulation, the LQG transitions amplitudes are defined  in terms of a sequence of 4d truncations, after fixing a relevant family of boundary states (say in ${{\cal H}_\Gamma}$). 
  Each truncation is defined by the choice of a 2-complex $\cal C$ having $\Gamma$ as boundary.   The spin foam amplitudes define a bra $\langle W_{\cal C}|$ on ${\cal R}_\Gamma$ \cite{Engle2007,Freidel:2007py,Kaminski:2009fm,Rovelli2015}.   The theory is well defined if refining the 2-complex the amplitude converges.  Numerical calculations give some partial positive indications that this can be the case exist, see for instance \cite{Frisoni2022}.

\section{Physical discreteness}% Ditt-invariance and limits
\label{Ditt-invariance}\label{discreteness}

It is important not to confuse the discreteness introduced by the various truncations used to define the theory (the graph of the spinnetworks, the two-complex of the spinfoams) with the physical Planck scale discreteness predicted by LQG. The first is only a theoretical tool, analog to the lattice of lattice QCD.   The second (absent in QCD) is a hard physical prediction of the theory, and the most characteristic feature of LQG. It is the analog of the discreteness of the spectra of the energy of the Hydrogen atom or non relativistic harmonic oscillator, or the discreteness of photons.  It is derived in the theory form the spectral analysis of the operators describing the geometry  \cite{Rovelli:1994ge}. It is compatible with the local Lorentz invariance of the theory \cite{Rovelli2003}.  It is this physical discreteness which is responsible for the ultraviolet finiteness of LQG and for the resolution of the singularities of general relativity \cite{Ashtekar:2008ay,Rovelli2013d}. 

The expression of the transition amplitudes as a spinfoam sum have much in common with a standard lattice discretization of a Feynman sum over histories \cite{Conrady:2008ea}, like the one that defines lattice QCD.  However, there is a crucial difference  \cite{Rovelli2011b}: in theories like lattice QCD the full quantum theory is recovered by sending the number of lattice sites to infinity as well {\em as the lattice spacing to zero}. Because of the underpinning diffeomorphism invariance, only the first of these limits (that is: refining the two complex) is required in LQG. See a detailed discussion in \cite{Rovelli2011b}.  

Here ``limit" must be understood in the sense of potential, not actual. 
What we do in physics is to compute transition amplitudes within approximations.  This is always done within a finite truncation, as we do in standard perturbative QFT and lattice QCD. Therefore the theory can be formally defined by  the continuous limit, but the actual theory to be used is always at arbitrary but finite truncation.\footnote{This observation may either be seen as a simple pragmatic consideration, or, perhaps, as a way to question the physical relevance of the actual limit theory \cite{Vidotto:2013oxa}.} 

A different question is how we recover {\em classical} general relativity. For this, we have to take both the continuum limit and a ``classical" limit: namely look at scales large with respect to the Planck scale. This is usually implemented as a large spin limit.  For a while, the LQG literature contained the wrong expectation (giving rise to an apparent ``flatness problem") that the classical limit could be taken before, and independently from, the continuum limit. This is not the case: the two limits must be taken together, see \cite{Han2014a,Rovelli2015,Han2017,Asante2020,Engle2022}.\footnote{Specifically: at fixed triangulation, the LQG amplitudes approximate sufficiently well the dynamics of discretized general relativity (Regge theory) only if the triangulation is sufficiently fine.}

\section{Three distinct notions of time}

In moving from non-relativistic physics to quantum gravity, the notion of time undergoes alterations similar to the notion of space.  However, the notion of time is more subtle than the notion of space, raising further issues. 

In the case of space, we observed that clarity is obtained by distinguishing the common relational notion of space, according to which objects are spatially located with respect to one another --a notion still in play in quantum gravity-- from the Newtonian notion of space, as a continuous metric manifold with an Euclidean geometry, which emerges only in approximations.  
%More precisely, continuity emerges in a $\hbar\to 0$ limit, the Lorentzian structure emerges on the tangent space and the euclidean structure in a $c\to\infty$ limit.  

In the case of time, the same distinction holds.  In everyday life we use a relational notion of time.  Time is just a counting of happenings in successions: for example the succession of days and years. It is a fact of nature that there are such successions of events.   Newtonian physics, on the other hand, postulates the existence of a physical time that is independent from any succession of events, and has a rich structure: it has a metric structure, it is the same all over the universe, defining a global simultaneity, and so on.  As well known, many features of such Newtonian time are approximations: they do not describe correctly the actual  temporal structure of reality.  There is no single canonical clock variable in the universe, and no global simultaneity, except in dramatic approximations like homogeneity and isotropy in cosmology.   

The absence of a single preferred time variable and the fact that no single variable has all the features typical of Newtonian time is just a fact of nature, and is the reason for the generalization of mechanics illustrate in section \ref{evoluzione}.  The formal structure illustrated in that section permits to define both the classical and the quantum dynamics coherently without having to specify a preferred time variable. As observed,  non relativistic physics describes evolution as change of the variables in time, relativistic physics describes evolution as change of the variables with respect to one another.

In the literature, a big deal had been made about the alleged existence of a  `problem of time' due to this absence of a canonical time.   The confusion in this issue stems from a mixing up two distinct questons:
\begin{enumerate} 
\item The first question is to understand how to describe dynamical evolution in a relativistic setting when there is no canonical time variable. \item The second problem is raised by our strong feeling that time ``flows" in a sense that makes it different from any other physical variable. In a Newtonian theory, we identify the flow of time with the change in the canonical variable $T$ of the Newtonian formalism.   But we cannot do so in a theory formulated in a way that does not select any time variable as special.  
\end{enumerate} 

The first of these two problems can be solved classically, as illustrated in Section \ref{evoluzione}: a dynamical theory does not require a specific time variable to be defined.  All predictions by a general relativistic theory can be obtained without specifying a canonical, or `special', time variable.  So, this question can be consistently answered. The quantities predicted by the theory are values of some variables when other variables have given values.   

A corresponding quantum formalism can also be defined, as shown in Section \ref{evoluzione}.  Alleged ``quantum" solutions of this same problem, such as the Page-Wootter construction \cite{Page:1983uc} are nothing else than this same solution expressed in the quantum domain. 

The second of the above problems, on the other hand, is rooted on a conceptual misunderstanding. We do experience a flow of time, of course.  To understand this experience we should look at our experience as it is in reality, and not assume that our experience reaches out directly to the deep and general structure of reality.  

What we experience is due to the specific and complex situation in which we are.  Not only our experience is in the Newtonian limit (so that we misinterpret aspects of this limit for universal features of nature), but it is also strongly marked by the fact that we access a small subset of the degrees of freedom of Nature, hence we experience macroscopic coarse grained variables that happen to have thermodynamic properties.  In particular, we happen to live in a universe with a strong entropy gradient, where the behaviour of these macroscopic observables has a marked irreversible character.  (We do not know why.  An hypothesis for the reason of this is in \cite{Rovelli:2015}, but this is irrelevant here.)  From the perspective of the fundamental theory, this fact is accidental.

A direct consequence of this fact is that our local present has abundant traces of the past \cite{Rovelli2020}, and past low entropy allows macroscopic histories to branch \cite{Rovelli2020a,Rovelli1956}. These facts determine the epistemic and the agential arrows of time, both aligned with the entropy gradient. These phenomena, not any preferred  fundamental  temporal variable, are principally responsible for the phenomenology of our experiential time. For an ample discussion of all this, see \cite{Rovelli2018}.  All this is very interesting, of course, but has nothing to do with quantum gravity.  

(Incidentally: a consistent thermodynamic and statistical theory of the classical gravitational field is still missing, let alone for the quantum one. This is one of the reasons of the confusion surrounding issues like black hole entropy. For hints and clumsy attempts in this directions, see for instance \cite{Rovelli2002c,Haggard2013,Chirco2016}.)

The notions of time that need to be distinguished in order to get clarity in quantum gravity are therefore three:
\begin{enumerate}
\item Relational time is the notion that allows us to say that two local events happen in a direct succession next to one another.  This is the analog of relational space.  This notion remains true in quantum gravity: we compute transition amplitudes for successions of local events.
\item Newtonian time is a quantity that is well defined only under numerous approximations are taken. Various features of Newtonian time are lost one after the other as approximations are undone. 
\item Experienced time includes a rich phenomenology that depends on the specific environment around us, especially the irreversibility due to the entropy gradient, and on the functioning of our brain and its functioning in terms of deliberations \cite{Ismael2022,Price2023}. 
\end{enumerate}

It is perhaps clarifying to distinguish a generic notion of ``change" from the specific concept of ``time".  By ``change" we may mean the most generic aspect of temporal contingency, in the following sense. We experience in the world that things can be in a certain way ``sometime" and different ``some other time".  This is a notion which is local (not global across the universe), not necessarily oriented, and does not require a single time variable to be described. That is, we describe the world  in terms of a certain number of quantities: $(A, B, C, ...)$  and the functional dependencies between these:  we can compile lists of observations $(A_1, B_1, C_1, ...) ...   (A_2, B_2, C_2, ...) ... $ giving us the values of these quantities ``changed".  Physics gives us equations that constrain these changes.  For instance the change in an oscillator is described by the two partial observables $(X,T)$  and their relation $f(X,T) =X - A \sin (\omega T - \phi) = 0$, where $A$ and $\phi$ are constants.  In this particular case, we can recognize ``$T$" as what we usually call time, but in general relativity there is no such easy recognition, and in general none of the variables  $(A, B, C, ...)$ have all the qualities we ascribe to time in Newtonian physics.   By ``time" we indicate a particular variable among those describing the world, that has a particular list of properties (for instance it is monotonic along the change) and in the approximate description of the world obtained in the non-quantum, non-relativistic limit is associated with the quantity measured by our clocks and with our experiential time: the sense of passing we have in our brain.  So, change and time are  different.   The first is part of the conceptualization in quantum gravity, the second is not.

\section{Conclusion}

A large number of important technical issues are open in quantum gravity\footnote{For instance, the  infrared ``bubble" divergences \cite{Riello2013,Frisoni2021,Dona2022,Han2021a}.}, not to mention the persistent lack of direct empirical support.  But LQG has at its disposal not only a powerful mathematical formalism that represents a tentative theory of gravity, but also a coherent conceptual picture within which a possible understanding of quantum spacetime can be framed. 

The relational notions of space and time that are familiar from our common experience remain useful in quantum gravity: events can be ``next" to one another spatially and ``next" to one another temporally.   Not so the structure of a general  relativistic spacetime, which only emerges in approximations.  A general covariant formalism for dynamics is well defined and clear: it is based on the notion of partial observables: quantities that  can be measured but in general cannot be predicted by themselves.  The dynamical theory gives the correlations between these, both in the classical and quantum domains.  The observable quantities in quantum gravity are the same as those of general relativity: in principle, any measurement in relativistic gravitational physics is also a measurement in quantum gravity. (Any measurement in relativistic gravitational physics can be represented as performed across a 3d surface in the form described above.) 

The entire theory has a strong relational character: localization in space and time is relational.  Measurements imply relations between spacetime regions.  Evolution is given as relative evolution. Quantum states are interpreted as relative states in the relational interpretation of quantum theory.  (The entropy gradient as well could be a perspectival phenomenon  \cite{Rovelli:2015}.) This deeply relational aspect of reality, that comes both from general relativity and from quantum mechanics, and that merges so naturally in quantum gravity, is perhaps the deepest insight that quantum gravity is offering into the nature or reality \cite{Vidotto:2022aoz}. 

\vskip1cm
%\pagebreak 

\noindent
\textit{Acknowledgments --~~} Sincere thanks to Emily Adlam and Pascal Rodriguez for their extensive and thoughtful comments, and to Muxin Han, Jared Wogan and Sofie Reid for comments on the draft of the manuscript.

This work was supported by the John Templeton Foundation  Grant  No.62312 \emph{``Quantum Information Structure of Spacetime''} (QISS). ~~ FV's work is supported by the Canada Research Chairs Program and by the Natural Science and Engineering Council of Canada (NSERC) through the Discovery Grant \emph{"Loop Quantum Gravity: from Computation to Phenomenology"}.
%We acknowledge the Anishinaabek, Haudenosaunee, L\=unaap\'eewak and Attawandaron peoples, on whose traditional lands Western University is located.%

%\bibliographystyle{utcaps}
%\bibliography{library,main}

%\end{document} 

\providecommand{\href}[2]{#2}\begingroup\raggedright\endgroup

\end{document}